\documentclass[a4paper,11pt]{article}%
\usepackage{amssymb,amsmath,amsfonts}
\usepackage{amsmath}
\usepackage{amsfonts}
\usepackage{amssymb}
\usepackage{graphicx}%
\providecommand{\U}[1]{\protect\rule{.1in}{.1in}}
\newtheorem{theorem}{Theorem}

\newtheorem{idea memo}[theorem]{Idea Memo}

\newenvironment{proof}[1][Proof]{\textbf{#1.} }{\ \rule{0.5em}{0.5em}}
\begin{document}

\title{{\LARGE No Zero Divisor for Wick Product in $(S)^{\ast}$}}
\author{Takahiro HASEBE, Izumi OJIMA and Hayato SAIGO\\RIMS, Kyoto University, Kyoto 606-8502, Japan}
\date{}
\maketitle

\begin{abstract}
In White Noise Analysis (WNA), various random quantities are analyzed as
elements of $(S)^{\ast}$, the space of Hida distributions ([1]). Hida
distributions are generalized functions of white noise, which is to be
naturally viewed as the derivative of the Brownian motion. On $(S)^{\ast}$,
the Wick product is defined in terms of the $\mathcal{S}$-transform. We have
found such a remarkable property that the Wick product has no zero divisors
among Hida distributions. This result is a WNA version of Titchmarsh's theorem
and is expected to play fundamental roles in developing the \textquotedblleft
operational calculus\textquotedblright\ in WNA along the line of
Mikusi\'{n}ski's version for solving differential equations.

\end{abstract}


\section{Introduction}

In the present notes, we verify one of the most essential algebraic properties
of the so-called U-functionals consisting of the $\mathcal{S}$-transforms of
the Hida distributions in the context of white noise analysis (WNA, for
short): they constitute an \textit{integral domain}, that is, a ring with no
zero divisors. For this purpose, we start from the Gel'fand triple,
\[
S(\mathbb{R})\subset L^{2}(\mathbb{R})\subset S^{\prime}(\mathbb{R}),
\]
together with the Gauss measure $\mu$ on $S^{\prime}(\mathbb{R})$ determined
through the Bochner-Minlos theorem by its characteristic function:
\[
\int_{S^{\prime}(\mathbb{R})}e^{i\langle x,f\rangle}d\mu(x)=e^{-\frac{1}%
{2}\langle f,f\rangle}.
\]
As usual, $S(\mathbb{R})$ and $S^{\prime}(\mathbb{R})$ here denote the spaces
of rapidly decreasing Schwartz test functions and of Schwartz tempered
distributions, respectively. $(S^{\prime}(\mathbb{R}),\mu)$ is the space of
white noise.

Extending test functions from $S(\mathbb{R})$ to $L^{2}(\mathbb{R})$, we
obtain Wiener's Brownian motion in such a form as%
\[
B_{t}:=\langle x,1_{[0,t]}\rangle.
\]
Then a \textquotedblleft higher Gel'fand triple\textquotedblright\ can be
constructed as follows [1]:
\[
(S)\subset(L^{2}):=L^{2}(S^{\prime}(\mathbb{R}),\mu)\subset(S)^{\ast},
\]
where the elements of $(S)$ and $(S)^{\ast}$ are called, respectively,
\textquotedblleft white noise test functionals\textquotedblright\ and
\textquotedblleft generalized white noise functionals\textquotedblright, the
latter of which are also called \textquotedblleft Hida
distributions\textquotedblright.

\section{\textbf{Definitions of} $\mathcal{S}$\textbf{- and} $\mathcal{T}%
$\textbf{-transforms}}

In constructing the framework of calculus on these spaces, the $\mathcal{S}%
$-transform [2] defined by
\[
\mathcal{S}:(L^{2})\ni\phi\mapsto(\mathcal{S}\phi)(\xi):=\int_{S^{\prime
}(\mathbb{R})}\phi(x+\xi)d\mu(x)\text{ \ \ \ for }\xi\in S(\mathbb{R}),
\]
plays crucial roles, especially in our context aiming at an
infinite-dimensional version of \textquotedblleft operational
calculus\textquotedblright. The domain $(L^{2})$\ of $\mathcal{S}$ can be
extended to $(S)^{\ast}$. Similarly, the $\mathcal{T}$-transform is introduced
by
\[
\mathcal{T}:(L^{2})\ni\phi\mapsto(\mathcal{T}\phi)(\xi):=\int_{S^{\prime
}(\mathbb{R})}\phi(x)e^{i\langle x,\xi\rangle}d\mu(x)\text{ \ \ \ for }\xi\in
S(\mathbb{R}),
\]
and its domain $(L^{2})$ is also extended to $(S)^{\ast}$. Both $\mathcal{S} $
and $\mathcal{T}$ are injective.

\section{\textbf{CCR structure of Hida derivatives} \textbf{and a new version
of quantum decomposition}}

We can define Hida derivative $\partial_{t}:(S)\longrightarrow(S)$ by%

\[
\partial_{t}:\phi(x)\mapsto\mathcal{S}^{-1}\left\{  \dfrac{\delta}{\delta
\xi(t)}(\mathcal{S}\phi)(\xi)\right\}  (x),
\]
which satisfies the following equality for all $\phi\in(S)$:
\[
\partial_{t}\phi(x)=D_{\delta_{t}}\phi(x):=\lim_{\varepsilon\rightarrow0}%
\frac{\phi(x+\varepsilon\delta_{t})-\phi(x)}{\varepsilon}.
\]
The adjoint $\partial_{t}^{\ast}:(S)^{\ast}\longrightarrow(S)^{\ast}$ of
$\partial_{t}$ can be identified with the map:
\[
\partial_{t}^{^{\ast}}:\phi(x)\longmapsto\mathcal{S}^{-1}\left\{
\xi(t)(\mathcal{S}\phi)(\xi)\right\}  (x).
\]
With the notation $x_{t}:=\partial_{t}^{^{\ast}}1\in(S)^{\ast}$ for the image
of the identity functional $1\in(S)^{\ast}$ under $\partial_{t}^{^{\ast}}$, we
have the relation $x(t)=\dfrac{d}{dt}B(t)$ valid for the $C^{\infty}$-mappings
$x(t):t\mapsto x_{t}$ and $B(t):t\mapsto B_{t}$, and hence, $x_{t}$ is called
\textquotedblleft the white noise at $t$\textquotedblright.

It is important that $\partial_{t}$ and $\partial_{t}^{\ast}$ satisfy the
canonical commutation relations. Moreover, they satisfy the relation%
\[
x_{t}\cdot\text{ }=\partial_{t}+\partial_{t}^{^{\ast}},
\]
as operators $(S)\longrightarrow(S)^{\ast}$. Here an operator $x_{t}\cdot$ on
$(S)$ to multiply the random variable $x_{t}$ is defined by
\[
\langle\langle x_{t}\cdot y,\phi\rangle\rangle=\langle\langle x_{t},y\cdot
\phi\rangle\rangle,
\]
in terms of the pointwise product $y\cdot\phi$ of $y$ and $\phi$ and of the
duality coupling $\langle\langle\cdot,\cdot\rangle\rangle$ between $(S)^{\ast
}$ and $(S)$. In reference to the \textquotedblleft finite-dimensional analog
of $\mathcal{S}$-transform\textquotedblright\ given with $d\mu_{1}%
(x):=e^{-\frac{x^{2}}{2}}dx/\sqrt{2\pi}$ by%

\[
\mathcal{S}_{1}:\phi(x)\mapsto(\mathcal{S}_{1}\phi)(\xi):=\int_{\mathbb{R}%
}\phi(x+\xi)d\mu_{1}(x),
\]
it can be understood that the above definitions and properties are quite
natural: under $\mathcal{S}_{1}$, the differential operator $\dfrac{d}{dx}$ is
transformed into $\dfrac{d}{d\xi}$ and $x-\dfrac{d}{dx}$ into $\xi$,
respectively. Note here that $x-\dfrac{d}{dx}$ is the adjoint of $\dfrac
{d}{dx}$ with respect to the normal Gauss measure $\mu_{1}$ owing to the relation:%

\[
\int_{\mathbb{R}}\left\{  \left(  x-\frac{d}{dx}\right)  f\right\}
ge^{-\frac{x^{2}}{2}}dx=\int_{\mathbb{R}}f\left\{  \frac{d}{dx}g\right\}
e^{-\frac{x^{2}}{2}}dx.
\]
Namely, the differential operators and their adjoints with respect to a
reference measure given by the Gaussian one, are not commutative but satisfy
canonical commutation relations. In sharp contrast, $\dfrac{d}{dx}$ and its
adjoint $-\dfrac{d}{dx}$ with respect to the Lebesgue measure are mutually
commutative. Moreover, $x-\dfrac{d}{dx}$ can also be viewed as
\textquotedblleft a renormalized form of $x$\textquotedblright\ due to the
non-Haar nature of the Gaussian measure. For instance, $(x-\dfrac{d}{dx}%
)^{n}1$ is nothing but the $n$-th Hermite polynomial, which can be seen as
\textquotedblleft the Gaussian version of $x^{n}$\textquotedblright. In this
way, the $\mathcal{S}$-transform naturally extends the algebraic structure
found in finite dimensional situations to the infinite dimensional ones.

\section{\textbf{Absence of zero-divisors }}

Using the $\mathcal{S}$-transform again, we can even introduce
\textquotedblleft the product of Hida distributions\textquotedblright. The
theorem below plays a key role.

\begin{theorem}
[{Potthoff and Streit [3]}]A complex-valued functional $F$ on $S_{c}%
(\mathbb{R})$ is the $\mathcal{S}$-($\mathcal{T}$-)transform of a Hida
distribution if and only if it satisfies the following conditions:\newline(i)
for any $\xi,\eta\in S_{c}(\mathbb{R})$, the function $F(z\xi+\eta)$ is an
entire function of $z\in\mathbb{C}$, \newline(ii) there exist constants
$K,a,p\geq0$ such that
\[
|F(z\xi)|\leq Ke^{{a|z|^{2}{{\Vert{H^{p}}\xi\Vert}_{2}}^{2}}}.
\]
(Here $H$ denotes the Hamiltonian of a harmonic oscillator with the spectrum
$\sigma=\left\{  2k+2|k=0,1,2,3,\cdot\cdot\cdot\right\}  $.)
\end{theorem}

Functionals satisfying the above conditions (i),(ii) are called U-functionals.
It can be proved that U-functionals form an algebra under the pointwise
product. As $\mathcal{S}$-($\mathcal{T}$-)transforms are injective, we can
define (two types of) products on $(S)^{\ast}$:%

\begin{align*}
a\diamond b:=  &  \hspace{3pt}\mathcal{S}^{-1}(\mathcal{S}a\cdot
\mathcal{S}b),\\
a\ast b:=  &  \hspace{3pt}\mathcal{T}^{-1}(\mathcal{T}a\cdot\mathcal{T}b),
\end{align*}
where $\diamond$ and $\ast$ are called, respectively, \textquotedblleft Wick
product\textquotedblright\ and \textquotedblleft convolution\textquotedblright%
\ with $\mathbf{1}$ and $\mathbf{\delta_{0}}$ (white noise delta-functional)
as their units. From the characterization theorem, we can derive the
fundamental property shared by these products:

\begin{theorem}
The products, $\diamond$ and $\ast$, have no zero divisor among Hida distributions.
\end{theorem}

\begin{proof}
Since $\mathcal{S}$- and $\mathcal{T}$-transforms are isomorphic, it suffices
to prove that there are no zero divisors among U-funtionals under the
pointwise product. We start from the equality $U_{1}U_{2}=0$ for two
U-functionals $U_{1}$ and $U_{2}$. This means $U_{1}(f)U_{2}(f)=0$ for all
$f\in S(\mathbb{R})$, and we can assume that $U_{1}$ is not identically zero,
which implies the existence of $f$ such that $U_{1}(f)\neq0$. By the condition
(i) in the above definition of U-functionals, we can take a sufficiently small
$\lambda$ for each $g$ so that $U_{1}(f+\lambda g)\neq0$ is valid, and hence,
$U_{2}(f+\lambda g)=0$. Because of the analyticity expressed in the condition
(i), this implies the equality $U_{2}(f+\lambda g)=0$ for all $\lambda$. Since
$g$ is arbitrary, $U_{2}$ vanishes identically: $U_{2}=0$.
\end{proof}

\section{\textbf{Summary and prospect}}

We summarize our results into such a basic scheme of the algebraic structures
inherent to WNA: the space $(S)^{\ast}$ of Hida distributions is mapped by the
$\mathcal{S}$- and $\mathcal{T}$-transforms isomorphically into the space of
U-functionals. In terms of these transforms, $(S)^{\ast}$ is equipped with two
kinds of commutative products, $\diamond$ and $\ast$, and is acted upon by the
CCR\ algebra consisting of Hida derivatives $\partial_{t}$, $\partial
_{t}^{^{\ast}}$ which provides the basic random variables $x_{t}\in(S)^{\ast}$
with its quantum decomposition, $x_{t}\cdot$ $=\partial_{t}+\partial
_{t}^{^{\ast}}$. Finally $(S)^{\ast}$ as well as its image under the
$\mathcal{S}$- and $\mathcal{T}$-transforms constitutes an integral domain
without zero divisors. We note that the validity of this result heavily relies
upon the complex-analytic structure involved in the framework of WNA.

To take advantage of the above picture, we recall here that Titchmarsh's
theorem guarantees the absence of zero divisors with respect to the
convolution product among such functions as continuous on $[0,\infty]$ and as
vanishing on $(-\infty,0)$. Heaviside's operational calculus has been
formulated by Mikusi\'{n}ski [4] in the use of the usual convolution product.
Since the above proposition guarantees the validity of the similar result, we
can naturally expect that it will allow us to develop a kind of
infinite-dimensional version of \textquotedblleft operational
calculus\textquotedblright\ in the context of WNA. As a generalization of the
It\^{o} integral, the Hitsuda-Skorohod integral can be introduced in terms of
$\partial_{t}^{\ast}f=x_{t}\diamond f$. Since $(S)^{\ast}$ has no zero
divisors, we can introduce the Wick-inverse of Hida distributions by
constructing the quotient field of $(S)^{\ast}$. They should play crucial
roles, for instance, in solving stochastic partial differential equations and
in analyzing singularities associated with many distribution-theoretical
aspects inherent to WNA [5].

The authors would like to express their sincere thanks to Prof.\ T.\ Hida for
his encouragement and interest in our joint project on the duality aspects in
WNA. They are very grateful to Messrs.\ R.\ Harada, H.\ Ando and K.\ Nishimura
for valuable discussions.

\end{document}